\documentclass[10pt,letterpaper,twocolumn]{article} 

\usepackage{ol2}
\usepackage[draft]{hyperref}
\usepackage{amsmath}

\begin{document}

\twocolumn[

\title{Coherent perfect absorption mediated anomalous reflection and refraction}


\author{Shourya Dutta-Gupta,$^{1}$ Rahul Deshmukh,$^{2}$ Achanta Venu Gopal,$^{3}$ Olivier J. F. Martin,$^{1}$ and S. Dutta Gupta$^{2,*}$}

\address{
$^1$Nanophotonics and Metrology Laboratory, Swiss Federal Insitute of Technology Lausanne, Lausanne CH-1015, Switzerland\\
$^2$School of Physics, University of Hyderabad, Hyderabad-500046, India\\
$^3$DCMPMS, Tata Institute of Fundamental Research, Mumbai-400005, India\\
$^*$Corresponding author: sdghyderabad@gmail.com
}

\begin{abstract}
We demonstrate bending of light on the same side of the normal in a free standing corrugated metal film under bi-directional illumination. Coherent perfect absorption (CPA) is exploited to suppress the specular zeroth order leading to effective back-bending of light into the `-1' order, while the `+1' order is resonant with the surface mode. The effect is shown to be phase sensitive  yielding CPA and superscattering in the same geometry.
\end{abstract}

]

\begin{figure}[h]
\centerline{\includegraphics[width=5cm]{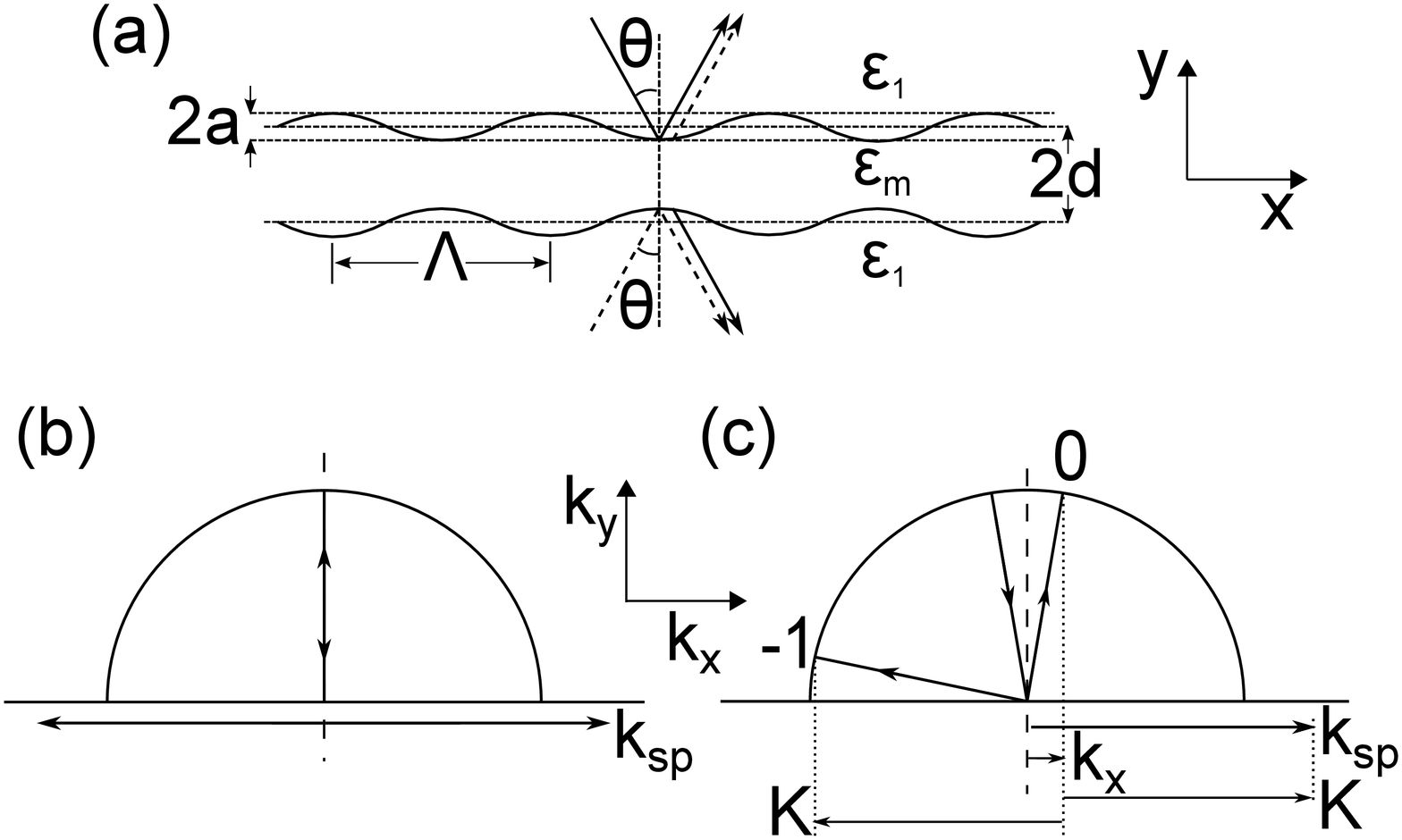}}
\caption{\label{fig1}(a) Schematics of the corrugated metal film and the illumination geometry (only the specular diffracted orders are shown). (b) Excitation of  SP's under normal incidence of light in reciprocal space. (\rm{c}) Excitation of $0$ and $-1$ harmonics as propagating modes for small angle incidence. Figures are not drawn to scale.}
\end{figure}

\noindent 
In recent years there has been a great deal of interest in perfect absorption \cite{graphene, audin, yoon, pu} motivated mostly by potential applications in light harvesting \cite{harvesting}. Many of such investigations focus on metallic gratings, which are known for their ability for near perfect absorption both in resonant and nonresonant regimes \cite{popov}. On the other hand a great deal of research has gone into critical coupling (CC) and coherent perfect absorption (CPA), whereby all the incident coherent light energy can be coupled into the structure leading to null scattering (simultaneous zero reflection and transmission) \cite{sdgol, subimal1, cao, wan, sdgopex, longhi, klimov}. Since the physical basis of both CC and CPA is destructive interference \cite{sdgol, sdgopex}, it is easy to guess the other extreme, namely, superscattering (SS), which results as a consequence of constructive interference. Similar ideas were exploited as early as in 1966 by Frederic \cite{frederick}, who showed that one can tune a grating from near-perfect absorption to superscattering  by bringing in an additional coherent light and controlling its phase. In this letter we explore analogous possibilities with a free standing corrugated metal film (see Fig. \ref{fig1}(a)) illuminated from both sides at the same angles of incidence. The illumination geometry is quite analogous to the one used earlier \cite{sdgopex} and the specular zeroth order light (both reflected and transmitted) can be suppressed for suitable system parameters. It is clear that for normal incidence and for suitable parameter (grating vector $K=2\pi /\Lambda$ resonant with surface plasmon (SP) wavevector $k_{sp}$) the $\pm 1$ orders can be excited corresponding to the SP's in both positive and negative $x$ directions (see Fig. \ref{fig1}(b)). Breaking the right-left spatial symmetry by tilting the incident light at small angles allows SP's only in the +1 order, while the -1 order can become propagating (see Fig. \ref{fig1}(\rm{c})). Coupled with the CPA assisted suppression of the specular zeroth order, this manifests itself as counterintuitive scattering of light on the same side of the normal. We show that in each side the $-1$ order scattered light intensity can be as high as 45 \% of the incident light. Recall that nearly half the energy goes in the excitation of the surface mode.
\par
Free standing metal films have been studied both theoretically and experimentally \cite{sdgprb,inagagi}. It was shown that for shallow gratings perturbative method like Rayleigh expansion is well suited and can lead to excellent agreements with the experimental results. For sufficiently thin films splitting into the long range (LR) and short range (SR) SP's was demonstrated. In this letter we use the same technique albeit with illumination by plane waves from both sides (see Fig. \ref{fig1}(a)). We use one of the split modes to demonstrate CPA mediated anomalous scattering, though in principle, the other could have been used. We show that the relative phase between the two incident waves can play an important role changing the character of interference from destructive (CPA) to constructive (SS).
\par
Consider a corrugated metal film of dielectric constant $\varepsilon_m$ suspended in air (dielectric constant $\varepsilon_1$) as shown in Fig.1(a). The surface profiles are given by the following equations
\begin{equation}
\label{eq1}
y_\pm=\pm d \pm a \sin{Kx},
\end{equation}
where the $+$ ($-$) sign refers to the top (bottom) interface and $2d$ and $a$ give the width and the corrugation amplitude, respectively. An analogous system was studied extensively and we employ the tested Rayleigh expansion into the spatial harmonics to calculate the various diffracted orders (both propagating and evanescent) \cite{sdgprb}. Convergence was tested by increasing the number of harmonics and it was satisfactory for the grating parameters used in calculation. 
\begin{figure}[t]
\centerline{\includegraphics[width=4cm]{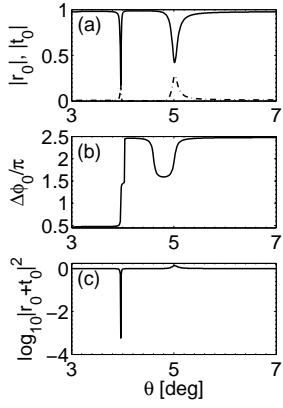}}
\caption{\label{fig2} (a) Absolute values of zeroth order reflected (solid line) and transmitted (dash-dotted line) amplitudes $|r_0|$ and $|t_0|$ , (b) phase difference $\Delta\phi_0/\pi$ and (\rm{c}) total scattered intensity on each side $log_{10}|r_0+t_0|^2$  as functions of angle of incidence for $\lambda=780~\rm{nm}$, $\Lambda=826~\rm{nm}$, $a=7.2~\rm{nm}$ and $d=27~\rm{nm}$, $\varepsilon_1=1.0$. The incident plane waves have a null phase delay.}
\end{figure}
\par
Significant development in nanofabrication technology has made it possible to realize sub-wavelength gratings with $K=k_{sp}$, whereby one can excite the SP's in the $\pm1$ orders for normal incidence (see Fig. \ref{fig1}(b)). CPA with bi-directional incidence can then suppress the specular order and thus all the incident energy will be converted into the surface plasmons in the $\pm x$ directions. Thus a SP in any given direction will have half of the incident plane wave energy. In our calculations we use the geometry shown in Fig.\ref{fig1}(\rm{c}), where one of the surface waves (in the $-x$ direction) is changed to a propagating mode, while the +1 order excites the SP.
\par
The following system parameters were chosen for most of the calculations: incident wavelength $\lambda=780~\rm{nm}$, $d=27~\rm{nm}$, $\varepsilon_1=1.0$. We used silver as the film material with $\varepsilon_m$ taken from the experimental work of Johnson and Christie \cite{johnson}. It is clear that such a thin film would enable the coupling of the SP's of the two interfaces leading to the LR and SR modes. Calculations were performed for two values of grating period $\Lambda$ so as to lead to two near-normal angles of incidence, satisfying plasmon resonance conditions at $+1$ diffraction order. The choice of parameters is not arbitrary, and they needed to be adjusted such that CPA condition is met for the zeroth specular order. Thus we used different values of the modulation depth $a$ for the same grating period $\Lambda$ in order to meet the CPA condition at the LR and SR modes.   Moreover, in order to demonstrate the phase sensitivity we introduced a delay in one of the incident beams. The delay will be assumed to be zero unless otherwise stated. It will be shown below how a change in this delay from zero to $\pi$ can change the character of scattering from SS to CPA. 
\begin{figure}[htb]
\centerline{\includegraphics[width=7cm]{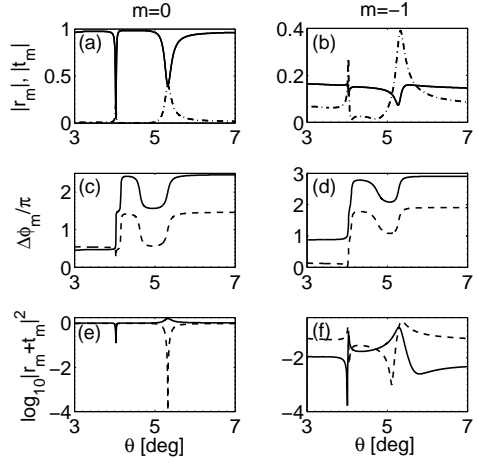}}
\caption{\label{fig3} Absolute values of reflected (solid line) and transmitted (dash-dotted line) amplitudes $|r_m|$ and $|t_m|$ (top row: (a) and (b)) , phase difference $\Delta\phi_m$ (middle row: (\rm{c}) and (d)) and total scattered intensity on one side $log_{10}|r_m+t_m|^2$ (bottom row: (e) and (f)), as functions of angle of incidence $\theta$ for the 0 order (left column) and -1 order (right column). The solid and dashed lines in (\rm{c})-(f) are for null and $\pi$ phase delay, respectively, between the incident waves. The other parameters are as in Fig.\ref{fig2}, except that now $a=8.9~\rm{nm}$.}
\end{figure}
\par
Let $r_m$ and $t_m$ represent the $m$-th amplitude reflection and transmission diffraction orders for unit amplitude p-polarized plane wave incidence. Let $\Delta \phi_m$ represent the phase difference (normalized to $\pi$) between $r_m$ and $t_m$. In the context of the zeroth order, the role of these quantities in CPA was discussed in detail in \cite{sdgopex} highlighting the importance of destructive interference, when $|r_0|=|t_0|$ and $\Delta \phi_0=\pm 1$. The results for CPA with the LR mode is shown in Fig.\ref{fig2}, where we have plotted  $|r_0|$ and $|t_0|$ (Fig.\ref{fig2}(a)), $\Delta \phi_0$ (Fig.\ref{fig2}(b)) and total scattered intensity on either side  $log_{10}|r_0+t_0|^2$ (Fig.\ref{fig2}(\rm{c})) for $\Lambda=826~\rm{nm}$ and $a=7.2~\rm{nm}$. One can easily see that at $\theta\sim4^\circ$ the CPA condition is met and one has near-null scattering at that angle.
\par
It may be noted that for a thin planar metal film the LR and SR modes differ by a phase difference of $\pi$, though perturbation of the surface in a corrugated film slightly offsets this difference. Otherwise,  same system parameters yielding CPA for the LR mode would lead to the SS at the SR mode. Because of the offset we had to pick slightly different system parameter, namely, $a=8.9 nm$ for the same grating period  $\Lambda=826 nm$ as in Fig.\ref{fig2} for demonstrating both SS and CPA with the SR mode. In particular the short range mode was picked for demonstrating CPA, since without interference it can never lead to null reflection. The results for  $|r_m|$ and $|t_m|$, $\Delta \phi_m$ (normalized to $\pi$),  and the log of the total scattered intensity $log_{10}|r_m+t_m|^2$ for two different orders, namely, $m=0$ and $m=-1$ are shown in Fig.\ref{fig3}. The results for $m=+1$ (not shown) confirms the SP-mediated local field enhancement.
\begin{figure}[htb]
\centerline{\includegraphics[width=7cm]{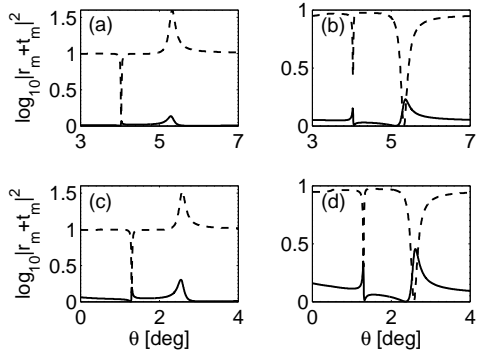}}
\caption{\label{fig4}  Total scattered intensity as a function of angle of incidence, for an initial phase delay of $0$ (left column) and $\pi$ (right column). The top and bottom rows are for $\Lambda_G=826~\rm{nm}$  and $\Lambda_G=787~\rm{nm}$, respectively. Dashed and solid lines in each panel are for $0$ and $-1$ harmonics, respectively. The other parameters are as in Fig.\ref{fig3}. }
\end{figure}
As can be seen from Fig.\ref{fig3}(a) that $|t_0|=|r_0|$ for an incident angle of $5.3^\circ$ implying that both CPA and CC can be realized provided that the phase difference is adjusted accordingly. As can be seen from Fig.\ref{fig3}(b) necessary conditions for CPA are met at the short range plasmon resonance for a phase delay of $\pi$ between the incident plane waves. In contrast incident waves with the same phase result in $\Delta\phi_0\sim2$ yielding superscattering. Both these features can be read from  
Fig.\ref{fig3}(\rm{c}) where we used a log plot to stress the null of scattering for CPA.  The right column in Fig.\ref{fig3} shows the results for the $m=-1$ diffraction order, which displays a dramatic change near the long range and short range modes. In fact, for a phase delay of $\pi$ between the input beams at $\theta=5.3^\circ$ (CPA for zeroth order) there can be significant scattering in the $-1$ order (Fig.\ref{fig2}(\rm{c})) at the SR resonance.
\par
In order to test our conjecture about the conversion of the evanescent SP to a propagating order under broken right-left symmetry with tilted incoming light, we reproduce the bottom panel of Fig.\ref{fig3} for two different values of grating period, namely, $\Lambda=826~\rm{nm}$ (top row in Fig.\ref{fig4}) and $787~\rm{nm}$ (bottom row) with parameters meeting the CPA/SS conditions for the SRSP. Dashed (solid) lines show the results for $m=0$ ($m=-1$) order. Left and right columns in Fig.\ref{fig4} correspond to phase delays $0~\rm{and}~\pi$, respectively, between the incident waves. Null phase delay corresponds to SS while a value of $\pi$ results in CPA at the SR mode resonance. Near-$\pi$ phase difference between the LR and SR modes lead to dips and peaks at the corresponding locations for the zeroth order (see for example, Fig.\ref{fig4}(a)). Note that the shorter grating period (nearer to the incident wavelength) corresponds to a smaller angle of excitation for the SRSP and is closer to the right-left symmetric structure. A comparison of Figs.\ref{fig4}(b) and (d) clearly reveal that for the shorter grating period one has larger total scattering in the $m=-1$ order and it can be as high as $45\%$. Thus near the SR resonance zeroth order gets suppressed due to CPA, while there is enhanced scattering in the only other propagating order, namely the $m=-1$ order. In other words light bends in the `wrong' direction on the same side of the normal.

\par
In conclusion, we have exploited CPA in a free standing corrugated metal film for near normal illumination from both sides to demonstrate apparent bending of light on the same side of the normal. The effect can have varied applications in nano antennae and in other areas of plasmonics.
\par
One of the authors (S Dutta Gupta) is thankful to Girish S Agarwal for fruitful discussions. This work was supported by the Swiss National Science Foundation (Grant No. CR23I2 130164).



\begin{thebibliography}{99}
\bibitem{graphene} S. Thongrattanasiri, F. H. L. Koppens, and F. J. G. de Abajo, ``Complete Optical Absorption in Periodically Patterned Graphene,''  \prl \textbf{108,} 047401 (2012).

\bibitem{audin} K. Aydin, V. E. Ferry, R. M. Briggs, and H. A. Atwater, ``Broadband polarization-independent resonant light absorption using ultrathin plasmonic super absorbers,'' Nature Communications \textbf{2}, 517 (2011).

\bibitem{yoon} J. W. Yoon, W. J. Park, K. J. Lee, S. H. Song, and R. Magnusson, ``Surface-plasmon mediated total absorption of light into silicon'', Opt. Exp.  \textbf{19,} 20673--20680 (2011).

\bibitem{pu} M. Pu, Q. Feng, M. Wang, C. Hu,C. Huang, X. Ma, Z. Zhao, C. Wang, and X. Luo, ``Ultrathin broadband nearly perfect absorber with symmetrical coherent illumination,'' Opt. Exp. \textbf{20,} 2246--2254 (2012).

\bibitem{harvesting} A. Polman and H. A. Atwater, ``Photonic design principles for ultrahigh-efficiency photovoltaics,'' Nature Materials \textbf{11,} 174--177 (2012).

\bibitem{popov} E. Popov and L. Tsonev,``Total Absorption of Light by Metallic Gratings and Energy Flow Distribution,'' Surface Science \textbf{230,} 290--294 (1990).

\bibitem{sdgol}~S. Dutta Gupta, ``Strong interaction mediated critical coupling at two distinct frequencies,'' Opt. Lett. \textbf{32,} 1483--1485 (2007). See also the references therein.

\bibitem{subimal1} S.~Deb, S.~Dutta Gupta, J.~Banerji and S.~Dutta Gupta, ``Critical coupling at oblique incidence,'' J. Opt. A : Pure and Appl. Opt. \textbf{9,} 555-559 (2007).

\bibitem{cao} ~Y.~D. Chong, L.~Ge, H.~Cao, and A.~D.~Stone, ``Coherent Perfect Absorbers: Time-Reversed Lasers,'' \prl \textbf{105,} 053901 (2010).

\bibitem{wan} ~W. Wan, Y.~D.~Chong, L.~Ge, H.~Noh, A.~D.~Stone and H.~Cao, ``Time-Reversed Lasing and Interferometric Control of Absorption,'' Science \textbf{331,} 889--892 (2011).

\bibitem{sdgopex} Dutta-Gupta,~S., O.~J.~F.~Martin, S.~Dutta~Gupta and G.~S.~Agarwal, ``Controllable coherent perfect absorption using a composite film,'' Opt. Exp. \textbf{20,} 1330--1336 (2012).

\bibitem{longhi} S. Longhi and G. Della Valle, ``Coherent perfect absorbers for transient, periodic, or chaotic optical fields: Time-reversed lasers beyond threshold,'' \prl \textbf{85,} 053838 (2012).

\bibitem{klimov} S. Longhi and G. Della Valle, ``Coherent perfect nanoabsorbers based on negative refraction,'' Opt. Exp. \textbf{20,} A13071 (2012).

\bibitem{frederick}
Frederick C. Evering, Jr., Artificial Diffraction Anomalies for Gratings of Rectangular Profile, \ao \textbf{5,} 1313--1317 (1966).


\bibitem{johnson} P. B. Johnson and R. W. Christy, ``Optical Constants of the Noble Metals, '' \prb \textbf{6,} 4370--4379 (1972)

\bibitem{sdgprb} S. Dutta Gupta, G. V. Varada, and G. S. Agarwal, ``Surface plasmons in two-sided corrugated thin films, '' \prb \textbf{36,} 6331--6335 (1987)

\bibitem{inagagi} T. Inagaki, M. Motosuga, E. T. Arakawa and J. P. Goudonnet, ``Coupled surface plasmons in periodically corrugated thin silver films,'' \prb \textbf{32,} 2548--2550 (1985).

\end{thebibliography}

\newpage
\end{document}